
%
%

\input psfig.tex
\magnification=\magstep0                                                  %
\font\titlefont=cmr10 scaled\magstep3                 %
\newbox\leftpage \newdimen\fullhsize \newdimen\hstitle \newdimen\hsbody   %
\hoffset=0.0truein \voffset=0.20truein \hsbody=\hsize \hstitle=\hsize     %
\tolerance=1000\hfuzz=2pt \baselineskip=20pt plus 4pt minus 2pt           %
\global\newcount\secno \global\secno=0 \global\newcount\meqno             %
\def\newsec#1{\global\advance\secno by1 \xdef\secsym{\ifcase\secno
\or I\or II\or III\or IV\or V\or VI\or VII\or VIII\or IX\or X\fi }
\global\meqno=1\bigbreak\bigskip\noindent{\bf \secsym. #1}
\par\nobreak\medskip\nobreak}                                             %
\def\appendix#1#2{\xdef\secsym{\hbox{#1}}                                 %
\global\meqno=1 \bigbreak\bigskip \noindent{\bf Appendix #1. #2}          %
\par\nobreak\medskip\nobreak}                                             %
\def\eqn#1#2{\xdef #1{({\rm\secsym}.\the\meqno)}\global\advance\meqno by1 %
$$#2\eqno#1$$}                                                            %
\global\newcount\ftno \global\ftno=1                                    %
\def\foot#1#2{{\baselineskip=14pt plus 1pt\footnote{$^{#1}$}{#2}}   %
\global\advance\ftno by1}                                               %
\global\newcount\refno \global\refno=1 \newwrite\rfile                    %
\def\ref#1#2{{$^{\the\refno}$}\nref#1{#2}}
\def\nref#1#2{\xdef#1{$^{\the\refno}$ }
\ifnum\refno=1\immediate\openout\rfile=refs.tmp\fi
\immediate\write\rfile{\noexpand\item{[\the\refno]\ }#2}
\global\advance\refno by1}                                                %
\def\c{$^,$}                                                  %

\def\fig#1#2{\medskip\item{Fig.~#1:  }#2}                                 %
\def\tab#1#2{\medskip\item{Table~#1:  }#2}
\def\del{\partial} 
\def\tiny{\scriptscriptstyle\rm\!}
\def\kf{k_{\, \tiny F}}
\def\qf{q_{\, \tiny F}}
\def\meas{\! {{dp d\omega} \over (2\pi)^2} \,}
\def\nup{\psi_{\uparrow}^{\dagger}\psi_{\uparrow}}
\def\ndown{\psi_{\downarrow}^{\dagger}\psi_{\downarrow}}

\def\pbyo{p \times 1}
\def\xiup{x_{\uparrow,i}}
\def\xjup{x_{\uparrow,j}}
\def\xidn{x_{\downarrow,i}}
\def\xjdn{x_{\downarrow,j}}
\def\uup{u_{\uparrow}}
\def\udn{u_{\downarrow}}
\def\tr{\theta_{\rho}}
\def\ts{\theta_{\sigma}}
\def\goes{\rightarrow}
\nopagenumbers \hsize=\hsbody \pageno=0 ~ \vfill                          %
{
\centerline{\titlefont  Roughening of Anisotropically Reconstructed
Surfaces}
\centerline{\titlefont and the Hubbard Model}
\vfill \centerline{{\sl Leon Balents}}
\centerline{Physics Department}
\centerline{Harvard University}
\centerline{Cambridge, MA 02138}
\vskip 1 truecm \centerline{{\sl Mehran Kardar}}
\centerline{Physics Department}                                        %
\centerline{Massachusetts Institute of Technology}                        %
\centerline{Cambridge, Massachusetts 02139}                               %
\vfill\centerline{\bf ABSTRACT}\nobreak\medskip\nobreak\par
{ We consider a model of a reconstructed crystal surface, first
considered by Villain and Vilfan\ref\rVillain{J. Villain and I.
Vilfan, Europhys. Lett. {\bf 12}, 523 (1990); and Europhys. Lett. {\bf
13}, 285 (1990).}\c\ref\rVVtwo{I. Vilfan and
J. Villain, Surf. Sci. {\bf 257}, 368 (1991).}\ for the gold (110)
surface, in which roughening occurs via the formation of
anisotropic steps traversing the entire length of the crystal. The
model is studied by a mapping to a spin--1/2 Fermion system in 1+1
dimensions, which, in the absence of islands, is precisely the Hubbard
model.  We consider a general $\pbyo$ reconstruction, in
the presence of inter--step interactions and closed islands.  Our
analysis predicts the existence of a new type of rough phase, with
incommensurate correlations in the reconstruction order parameter and
unusual momentum space singularities at a characteristic ``Fermi
momentum'' and its harmonics.  The new phase is analagous to the
Luttinger liquid of one--dimensional Fermions.  The general phase
structure is as follows: for $p=1$, the surface has only an ordinary
roughening transition.  For $p>2$, there is a flat ordered (FO),
a rough incommensurate (RI), and a flat incommensurate phase (FI).  The
FO--RI and FO--FI transitions are of the commensurate to incommensurate
type, and the FI--RI transition is in the Kosterlitz--Thouless (KT)
universality class.  For $p=2$, the FI phase is replaced by a flat
disordered phase (FD), and there may be a new rough disordered phase
(RD).  The FO--FD transition is now of Ising type, and the FD--RD and
RI--RD transitions are in the KT universality class.  There is also the
possibility of a direct RI--FD transition, in which the height--height
stiffness has a {\it nonuniversal} jump at the transition.

}}
\vfill\eject\footline={\hss\tenrm\folio\hss}                              %
\vskip .3in
\newsec{Introduction}

The behavior of the two dimensional surfaces of three dimensional
crystals is of great interest as a physical system which can exhibit a
variety of behaviors and is easily accessible experimentally.  In
addition, surface physics plays an important role in a large number of
technological processes.  One well studied surface phenomenon is the
roughening transition,\ref\rRough{ J.D. Weeks, in {\it Ordering in
Strongly Fluctuating Condensed Matter Systems,} edited by T. Riste,
(Plenum Press, New York 1980), p. 293. }\c\ref\rChuiweeks{S.T.  Chui and
J.D. Weeks, Phys. Rev. B {\bf 14}, 4798 (1976).}\ in which the flat nature of
a crystalline surface is destroyed at high temperatures.  Another
extremely common fate of surfaces is a reconstruction or restructuring
transition, in which the surface develops a different lattice
structure from the bulk of the material.  In some systems both types
of processes are present, and for such surfaces we find a new type of
rough phase.

The ordinary roughening transition corresponds to the proliferation of
more or less isotropic islands within which the surface height jumps
by one unit.  This transition is now well known in its conventional
form to be in the Kosterlitz--Thouless (KT) universality
class.\ref\rKT{J.M. Kosterlitz and D.J. Thouless, J. Phys. C {\bf
6}, 1181 (1973). }\ In such systems, the rough phase has
logarithmically growing height--height fluctuations, and is
characterized by a single ``stiffness'' parameter $K$, such that
\eqn\eStiffnorm{ \langle \left[ h(x) - h(0) \right]^2 \rangle \sim {2K
\over {\pi^2}}\log |x|,}
where $h(x)$ is the height of the surface at the point $x$.  The
theory of the roughening transition tells us that $K$ takes on a
universal value of $1$ at the transition.

Reconstruction transformations involve a change in the regular lattice
structure at the crystal surface.  Such a change is pictured in
Fig.[1] for the case of the Gold (110) plane, known as a $2 \times 1$
reconstruction, due to the fact that the ledges comprising the new
surface have sides of length two.\ref\rGoldrough{I.K. Robinson, E.
Vlieg, and K. Kern, Phys. Rev. Lett. {\bf 63}, 2578 (1989). }\ This
particular restructuring is anisotropic, in that this pattern extends
uniformly perpendicular to the ledges; many other surfaces have
isotropic reconstructions.\ref\rMochrie{D. Gibbs, G. Gr\"ubel, D.M.
Zehner, D.L. Abernathy, and S.G.J. Mochrie, Phys. Rev. Lett {\bf
67}, 3117 (1991). }\ The transition to the restructured state
involves symmetry breaking, since the crystal has a choice of possible
sublattices for the new surface pattern.  For the case of a $\pbyo$
reconstruction (the straightforward generalization of the Gold (110)
case above), the symmetry group relating the possible sublattices is
just a group of $2p$ discrete translations (see Fig.[2]).
An order parameter distinguishing these phases is
\eqn\eRoughorder{ R(x) \equiv \exp \left({2\pi i \over {2p}} \delta(x)
\right),} where $\delta(x) = 0,1,\ldots,2p-1$ is the discrete translation
distance specifying the sublattice at position $x$.  In this
anisotropic system, the unreconstructed phase can be created from the
reconstructed one by the formation of domain walls separating
regions with different values of $R(x)$.  The statistical mechanics of
these domain walls is analagous to that of the
commensurate to incommensurate transition (CIT).\ref\rHalperin{H.J.
Schultz, B.I.  Halperin, and C.L. Henley, Phys. Rev. B. {\bf 26}, 3797
(1982).}\c\ref\rCoppersmith{S.N. Coppersmith, D.S. Fisher, B.I.
Halperin, P.A. Lee, and W.F. Brinkman, Phys. Rev. B {\bf 25}, 394
(1982).}\ As in the CIT, there is algebraic decay of correlations in
the unreconstructed phase, i.e.
\eqn\eUnrecquasi{ \langle R(x)R^{*}(0) \rangle \sim  {1 \over
x^\eta}\cos\left( {{2\pi n x} \over p} \right), }
where $\eta$ is proportional to a ``stiffness'' parameter describing
the restructuring, and $n$ is the average density of domain walls.
Such algebraic decay is known as quasi--long--range
order, a common phenomenon in two dimensions.  True long--range order,
where $\langle R(x)\rangle \neq 0$, occurs in the reconstructed phase,
while a truly disordered phase may occur at higher temperatures when
loops are relevant.  In most isotropic systems, the transition will be
directly to a disordered phase.

Recently, there have been theoretical and experimental studies of
roughening of reconstructed crystal surfaces.  Early experimental
studies of the $2 \times 1$ reconstruction of the gold (110) surface
showed unusual behavior of correlation functions in the rough
phase.\rGoldrough\ More recently, this work has been refined and
indicates the presence of both an Ising and a roughening
transition.\ref\rGoldexpb{J.  Spr\"osser, B. Salanon, and J.
Lapujoulade, Europhys. Lett {\bf 16}, 283 (1991).}\ Experiments on the
$2 \times 1$ surface of Pt (110)\ref\rRVK{I.  K. Robinson, E. Vlieg, and
K. Kern, Phys. Rev. Lett. {\bf 63} 2578 (1989).}\ showed simultaneous
deconstruction and roughening, with critical exponents
consistent with Ising values.  There is also published work on a $3
\times 1$ reconstruction of gold\ref\rBirgeneau{G.A. Held, J.L.
Jordan--Sweet, P.M. Horn, A. Mak, and R.J. Birgeneau, Solid State
Commun. {\bf 72}, 37 (1989).},
in which a single transition
combining both roughening and deconstruction was observed.  In
response to the first of these papers, Villain and Vilfan (VV)
proposed a simple theoretical model based on step formation to explain
the experimental observations.\rVillain\c\rVVtwo  Though we will
concentrate on the picture introduced by VV, several other theoretical
papers have also attempted to explain the experimental observations.
Den Nijs\ref\rDenNijs{ M. den Nijs, Phys. Rev. Lett. {\bf
66}, 907 (1991).}\ performed finite size simulations on a
phenomenological lattice model (see sections II and VII for the
relationship with this paper), while Mazzeo {\it et. al.}\ref\rMazzeo{
G. Mazzeo, G. Jug, A. C. Levi, and E. Tossati, SISSA preprint
(2/92).}\ used microscopic ``glue model'' calculations to provide
parameters for their Monte Carlo simulations.

The VV model considers the possibility that, in some systems,
roughening may occur via the appearance of anisotropic `steps'
crossing the length of the crystal surface.\rVillain\ A step on this
surface consists of a modification of one of the ledges comprising the
reconstruction (see Fig.[3]).  In this case the geometry of the
surface introduces constraints that prohibit the formation of simple
islands, although more complicated ones are possible (see Fig.[4]).
Entropic considerations suggest that these complex loops may be
irrelevant in some regimes, so, following VV, they will be
ignored to a first approximation (and included later on).  In this
case, the model can be formulated as the physics of a system of
interacting directed lines, with the restriction that (assuming that
energy of steps of more than one elementary lattice height is
sufficiently large) two up or two down steps cannot cross.  VV
studied this model in the absence of any additional interactions, and
gave rough arguments concerning its behaviour in other
regimes.\rVillain\ They made some use of a well--known mapping from the
statistical mechanics of directed lines in two dimensions to	 the
quantum mechanics of Fermions in one dimension.\ref\rPokTal{V.L.
Pokrovsky and A.L. Talapov, Phys.  Rev. Lett. {\bf 42}, 65 (1979).}\
In this paper, we apply the full power of the Fermion mapping to the
problem.

Using a generalized version of the VV model, we demonstrate the
existence in many reconstructed systems of a new rough phase.  This
new phase bears much the same relationship to the conventional rough
phase that the ``Luttinger liquid'' state of one--dimensional Fermions
does to the conventional Fermi liquid.  Although the height--height
correlations remain logarithmic, the new rough phase is described by
additional parameters.  These parameters can be thought of as arising
from the presence of the reconstruction order parameter, which is also
critical in the new rough phase.  Although the two types of
correlations behave independently as in the ordinary rough or
incommensurate phase, they are in fact coupled.  In addition to the
logarithmic height fluctuations, the theory predicts continuously
varying singularities in various correlation functions near a
characteristic ``Fermi momentum'' and its harmonics.

In a real crystal, there are many interactions between different
steps.  As a function of the strength of these interactions and the
chemical potential for step formation, the system falls into several
phases, including the new rough phase described above.  For a general
$\pbyo$ reconstructed surface, the phase diagram depends upon whether
$p=1$, $p=2$, or $p > 2$.  For $p=1$, the system is always unstable to
loop formation, and there is only a single reconstruction transition
(see Fig.[5a]).  For $p>2$, the system is stable to loop
formation, and the phase diagram is given by Fig.[5b].
There is a flat ordered (FO) phase in which no steps are present, a
flat incommensurate (FI) phase in which up and down steps are bound,
but the reconstruction order parameter has floating correlations, and
the new rough (RI) phase, in which the reconstruction order parameter
also has floating correlations, but with a different periodicity.  The
FI--FO and RI--FO transitions are both in the CIT universality class,
while the RI--FI transition may be either KT or a modified form also
containing only essential singularities.  For $p = 2$ (see Fig.[5c]),
loops are relevant in the FI phase and irrelevant in the RI phase.
The FI phase is replaced by a flat disordered (or unreconstructed)
(FD) phase in which double loops (Fig.[4]) proliferate.  The RI--FO
transition remains of the CIT type, while the FD--FO transition is
Ising--like.  Between the FD and RI phases there may now appear a rough
disordered (RD) phase, depending upon the strong--coupling behavior of
the theory.  If this phase is present, the FD--RD transition will be of
the KT type, while the RD--RI transition will be in the KT$^*$ class,
where the $^*$ indicates that the universal jump at the transition is
in the {\it reconstruction} stiffness, and not in the corresponding
height variables.  If the RD phase is not present, the RI--FD
transition will be in the KT$^{*}$ universality class, with the height
stiffness undergoing a {\it non--universal} jump at the transition.

In section II we describe the non--interacting version of the model and
calculate thermodynamic quantities and correlation functions in this
limit.  In section III, we introduce a contact interaction between up
and down steps.  The resulting system is identical to the Hubbard model
of spin--1/2 Fermions with contact interactions.  The latter model is
exactly soluble by the Bethe Ansatz\ref\rLiebWu{E.H. Lieb and F.Y.
Wu, Phys. Rev. Lett. {\bf 20}, 1445 (1968).}, from which we derive
exact expressions for the free energy, heat capacity, and mean
density.  In section IV, a Fermion renormalization group procedure is
introduced to address the questions of singularities at small
interaction, and of correlation functions. We encounter a fixed line
which is studied in section V.  Here the ``Luttinger liquid''
description of Haldane\ref\rHaldane{ F.D.M.  Haldane, J. Phys. C
{\bf 14}, 2585 (1981).}\
allows us to understand the long
distance properties of correlation functions in the RI phase.
Section VI describes how these conclusions are modified when generic,
non--contact interactions between steps are allowed.  In section VII,
we study the stability to loop formation of the general $\pbyo$
system, and describe how the physics is altered when these loops are
relevant.  Our conclusions, and suggestions for experimental
observation of the new effects, are given in section VIII.  Details of
the renormalization group calculation, the Bosonization mapping, and
the relationship between our work and methods used for the CIT, are
given in appendices A, B, and C.  Appendix D contains a brief
discussion of the related Fermion mapping for the simpler case of
vicinal surfaces.

\newsec{Non--interacting Steps and Free Fermions}

On a reconstructed crystal surface, steps can form as ``defects'' in
the orderly pattern of reconstruction.  The surface geometry
corresponding to these steps is pictured for Gold (110) in Fig.[3].  A
combination of one up and one down step returns the surface to its
original height.  As noted by VV,\rVillain it is also possible
to assign a parity to each atom along the surface.  While the up/down
step combination returns the surface to its original height, the
parity is reversed.  It is thus impossible to form a closed loop of
raised surface, since one cannot connect the two different parity
regions around the loop.  Such a loop is pictured in Fig.[4a].  It
{\it is} possible, however, to form a loop using two up and two down
steps, since two reversals of parity are equivalent to no reversal
(See Fig.[4b]).  In general, for a $\pbyo$ reconstruction,
loops with $p$ up and $p$ down steps form the smallest allowed
islands.

As noted by den Nijs\rDenNijs, other types of defects also form on the
reconstructed surface.  In the simplest $2 \times 1$ case, there are
also ``wall'' defects, which change the surface sublattice but not the
height,  and steps of opposite ``chirality'' (den Nijs' terminology).
These steps consist of a $1 \times 1$ microfacet in contrast to the $3
\times 1$ microfacets shown in Fig.[3].  As a consequence, the
corresponding sublattice shift is {\it opposite} to that of the $3
\times 1$ step.  As the temperature is raised, the defect with the
smallest free energy cost will unbind first. den Nijs works in the
limit that the free energy difference between opposite chirality steps
is very small, which leads to a completely  isotropic theory.  We will
consider the opposite limit in which only a single chirality of step
is present, and the surface has the potential to remain anisotropic.

A simplified version of this problem has only a
single species of step.  This is the case for a miscut (vicinal)
surface in which a large number of up steps are necessary to
accomodate the misorientation.  This system
has been studied using the methods that we now generalize to
up/down steps.\ref\rVicinal{See, for example, B. Joos, T. L. Einstein
and N. C. Bartelt, Phys. Rev. B {\bf 43}, 8153 (1991); X.-S. Wang,
J.L. Goldberg, N.C.
Bartelt, T.L. Einstein, and E.D. Williams, Phys. Rev. Lett. {\bf
65}, 2430 (1990).}\
Each step is a directed line traveling along the $z$ axis,
described by a transverse coordinate $x(z)$.  The simplest continuum
expression for the partition function of a single line connecting the
points $(0,0)$ and $(x,z)$ is
\eqn\ePath{Z_1(x,z) = \int_{(0,0)}^{(x,z)} [dx(z)] \exp \left\{ -{1 \over
{2\gamma}}\int_{0}^{z} \! dz' \left( {dx(z') \over dz'} \right) ^{2}
\right\} ,}
where $\gamma$ is the inverse line tension (all higher order terms are
irrelevant). The partition function for the full system
of many steps is a product of the individual partition
functions, with additional coupling terms that account for their
interactions.  For low densities, there is a repulsive interaction
between steps of the same species that is effectively equivalent to a
non--crossing condition\ref
\rJV{J. Villain, in {\it Ordering in Strongly Fluctuating Condensed Matter
Systems}, edited by T. Riste (Plenum, New York, 1980), p. 221.}.
There are also additional interactions between up and down steps,
as well as non--local interactions between steps of the same species.

In the non--interacting version of the VV model, steps of double height
are forbidden, but up and down steps are allowed to cross without
penalty and are in all other ways free.  The non--crossing condition
for each species of steps is easily implemented by associating the
steps with the world lines of a system of Fermions.\rPokTal\ The
exclusion principle then prevents the lines from crossing, and the
imaginary time ($z$) Green's function
\eqn\eTransfer{Z = {\rm Tr} \,e^{- H T},}
is also the full partition function. $T$ is the size of the
system in the $z$ direction, and $H$ is the quantum
Hamiltonian
\eqn\eHam{ H = -{\gamma \over 2}\sum_{i}\left[{\del^{2} \over {\del
x_{\uparrow,i}}^2} + {\del^{2} \over {\del
x_{\downarrow,i}}^2} \right] + \sum_{i,j} \left[
V_{1}(\xiup - \xjup) + V_{1}(\xidn-\xjdn) +
V_{2}(\xiup -\xjdn) \right].}
Here we have included the potential terms that will be used in future
sections.  With only the single step height restriction, these terms
are not present.

For ease in later
sections, we now introduce a second quantized notation for the
quantum problem.  By standard methods of many body physics, the
Hamiltonian \eHam\ can be reexpressed as
\eqn\eHamsecquant{H = \sum_{\alpha}\int \!\!d\!x \,
\psi_{\alpha}^{\dagger}\! \left( {-{\gamma \over 2}
\partial_{x}^2 - \mu } \right) \! \psi_{\alpha} + \sum_{\alpha,\beta}\int
\!\!d\!x d\!y \,
n_{\alpha}(x) V_{\alpha\beta}(x-y) n_{\beta}(y), }
where $\alpha,\beta = \uparrow,\downarrow$, $\psi_{\alpha}$ and
$\psi^{\dagger}_{\alpha}$ are annihilation and creation fields for
species $\alpha$, and the local densities are expressed as
$n_{\alpha}(x) = \psi^{\dagger}_{\alpha}(x)\psi_{\alpha}(x)$ in terms
of the field operators.

The ground state energy determines the free energy, since the trace in
eq.\eTransfer\ is dominated by the state of lowest energy as $T \goes
\infty$. For a fixed step density, this is just a pair of filled Fermi seas.
The free energy density is thus given by
\eqn\eGroundfree{{F\over LT}={E_{0}\over L} = 2 \int_{-\kf}^{\kf} \! {dk \over
{2\pi}} \gamma k^2/2 = {\gamma \over {3\pi}}\kf^3 = {\pi^2 \over
3}\gamma n^3,}
where $\kf=n\pi$ is the Fermi momentum, and $n = n_{\uparrow} =
n_{\downarrow}$ is the density of each species of steps.  To
compute the grand canonical partition function, we must add a chemical
potential term, $-\mu n$, and minimize over the density.  At the
minimum,
\eqn\eDensity{\langle n \rangle = \left( {\mu \over {\pi^2\gamma}}
\right)^{1/2},}
and
\eqn\eFreefree{\langle f \rangle = -{2\mu^{3/2} \over
{3\pi\gamma^{1/2}}}}
for $\mu>0$, and both quantities are zero otherwise.

There are two types of correlation functions of interest, probing the
roughness and reconstruction respectively.  The first
measures the height--height fluctuations of the surface,
\eqn\eCorrdef{ C(x,z) \equiv \langle \left[ h(x,z) - h(0,0) \right]^2 \rangle,}
and is computed by expressing the height difference in terms of the
Fermion density operators.  For simplicity, we consider only the
equal $z$ correlations, $C(x,0)$.  Since the height jumps (drops) by one
unit each time an up (down) step is crossed, the
height difference is given by
\eqn\eHeightdiff{ h(x,z) - h(0,z) = \int_{0}^{x}\!\!dy \, S(y,z),}
where $S(y,z) \equiv n_{\uparrow}(y,z) - n_{\downarrow}(y,z)$ is the local
slope, corresponding to the local spin density in
the Fermion analogy.  In the quantum picture, the thermal average
of eq.\eCorrdef\ is simply a ground state expectation value.
For equal time correlations, we can use
Schr\"odinger picture operators, and drop the $z$ label.  In Fourier space,
\eqn\eCorrfourier{C(x,0) = \int \! {{dp dp'} \over (2\pi)^2} \,
f_{x}(p)f_x(p') \langle S(p)S(p') \rangle, }
where $f_{x}(p) \equiv (e^{ipx} -1)/(ip)$.

To evaluate eq.\eCorrfourier, we insert a complete set of states
between the two ``spin'' operators.  The first term in this
set produces only ground state expectation values, which vanish
for $S = n_{\uparrow} - n_{\downarrow}$.  The next term contains the
lowest excitations which are single particle/hole pairs.  All higher order
excited states will not contribute, since the density,
composed of a single pair of creation/annihilation operators,
can connect only to single pair excitations.  The two possible
species of pairs contribute equally, so that
\eqn\eCorrsplit{ \langle S(p)S(p') \rangle = 2 \int\! {{dk_p dk_h}
\over (2\pi)^2} \langle 0 | n_{\uparrow}(p) | k_p k_h \rangle \langle k_p k_h |
n_{\uparrow}(p') | 0 \rangle,} where $|k_h| < \kf$ and $|k_p|>\kf$.
The matrix elements simply yield delta functions restricting $p = k_h
- k_p$, and $p' = -p$ as expected from momentum conservation. Thus
\eqn\eCorrnice{C(x,0) = 4 \int\! {{dk_p dk_h} \over (2\pi)^2} { {1 -
\cos (k_h - k_p)x } \over (k_h - k_p)^2},}
where again the momentum range is restricted as in eq.\eCorrsplit\ .
The large $x$ behavior of $C(x,0)$ is therefore
\eqn\eCorrlarge{C(x,0) \sim {2 \over \pi^2} \log |x| .}

The reconstruction order parameter correlations can be computed in a
similar way.  Again, the simplest results are obtained for the
equal--$z$ correlations.  Starting with the expression for the order
parameter,
\eqn\eRlinesexp{R(x,z) = \exp \left[ {{\pi i} \over p}\int_{-\infty}^{x} \!\!
dy \, \bigg( n_{\uparrow}(x) + n_{\downarrow}(x)\bigg) \right],}
the roughening correlation function of interest becomes
\eqn\eRoughcorr{G(x,0) \equiv \langle R(x,z)R^{*}(0,z) \rangle =
\left\langle \exp \left[ {{\pi i} \over p} \int_{0}^{x} \! dy \, \bigg(
n_{\uparrow}(x) + n_{\downarrow}(x)\bigg) \right] \right\rangle. }
This is easily evaluated using the methods of the previous
calculation, resulting in
\eqn\eRoughcorrresult{ G(x,0) \sim x^{-1/p^2} \cos \left( {{2\pi n x}
\over p} \right), }
for large $x$.  This is an example of the so called ``floating'' or
``incommensurate'' ordering well known in the context of the
CIT.\rHalperin\c\rCoppersmith

In this special limit, roughening and deconstruction
occur simultaneously as $\mu \rightarrow 0$.  This suggests the
possibility of non--KT roughening transitions for anisotropic surfaces.
We investigate this possibility in more detail for the full interacting
problem in the remainder of the paper.

\newsec{Interacting Steps, the Hubbard Model, and the Bethe Ansatz}

%
%

To understand the effects of interactions between steps,
first consider the simple short--range limit which is exactly soluble.
In fact, inter--step forces on crystalline surfaces are generally
long--ranged, and such cases will be considered later in section VI.  The
locality assumption amounts to taking $V_{\alpha\beta}(x) =
c_{\alpha\beta}\delta(x)$ in eq.\eHamsecquant. Since double
height steps are forbidden (corresponding to the exclusion principle),
only the off--diagonal terms in $c_{\alpha\beta}$ survive.  Both terms
are equivalent, so that $c_{\uparrow\downarrow} =
c_{\downarrow\uparrow} \equiv c$, and
\eqn\eCont{ H = \sum_{\alpha}\int \!\!d\!x \, \psi_{\alpha}^{\dagger} \!
\left( {-\partial_{x}^2 - \mu } \right) \! \psi_{\alpha} + 2c\int
\!\!d\!x \,
\psi^{\dagger}_{\uparrow}  \psi^{\dagger}_{\downarrow}
\psi_{\downarrow} \psi_{\uparrow}, }
where a scale has been selected to fix the coefficient of the
kinetic term to one.  This Hamiltonian is identical to that of the
spin--1/2 Fermi gas with delta function interactions, which is the
continuum limit of the Hubbard model.\ref\rHubbard{V.J. Emery, in
{\it Highly Conducting One--Dimensional Solids,} edited by J.T.
Devreese {\it et. al.} (Plenum, New York, 1979), p. 327. }\
This theory has been the subject of much current interest due to
its possible relevance to the understanding of high temperature
superconductivity.\ref\rHiTc{See, for instance, P.W. Anderson, Phys.
Rev. Lett. {\bf 67}, 3844 (1991).}\ A great deal is known about the
model, and it would be very good if some of these predictions could be
tested in surface systems.

The exact solution of the Hubbard model was originally discovered by
Lieb and Wu\rLiebWu by Bethe Ansatz techniques.  They calculated all
thermodynamic quantities exactly in terms of a `spectral density
function' $\rho(p)$, which is the solution of a certain integral
equation.  Excitation energies are also obtainable as solutions of
other integral equations.
The continuum model is, of course, also exactly soluble by the same
methods as the Hubbard model, with the added advantage that the integral
equations are somewhat simplified.  On the attractive side ($c<0$), the
equation is
\eqn\eInteqnatt{2\pi\rho (p) = 2 - 2\int_{-\kf}^{\kf} K(p-p';2|c|) \rho(p')
dp',}
where $K(p;c) = c /(c^2 + p^2)$.  The density $n \equiv
N_{\uparrow}/L = N_{\downarrow}/L$ and energy $E$ are given by
\eqn\eEnmomatt{\eqalign{
n = & \int_{-\kf}^{\kf} \rho(p) dp, \cr
{E\over L} = & 2\int_{-\kf}^{\kf} (p^2 - c^2) \rho(p) dp. \cr}}
On the repulsive side, there are
two integral equations,
\eqn\eInteqnrep{\eqalign{
2\pi\rho (p) = & 1 + 2\int_{-\qf}^{\qf} K(p-p';c)\sigma(p')dp', \cr
2\pi\sigma (p) = & - 2\int_{-\qf}^{\qf} K(p-p';2c)\sigma(p')dp' +
2\int_{-\kf}^{\kf} K(p-p';c) \rho (p') dp' .\cr}}
In this case the Fermi momenta are fixed by the conditions
\eqn\eFermimom{\eqalign{
2n = & \int_{-\kf}^{\kf} \rho(p) dp, \cr
n = & \int_{-\qf}^{\qf} \sigma(p) dp, \cr
{E\over L} = & \int_{-\kf}^{\kf} p^2 \rho(p) dp . \cr}}

In principle, these equations allow a precise calculation of the
singularities in the free energy across all possible phase boundaries.
For strong coupling ($n/c \rightarrow 0$), it is possible to solve the
integral equations perturbatively.  The spectral density functions become
\eqn\eSpectral{\eqalign{
\rho_{\rm attr.}(p) = & {1 \over \pi} + O(n/c), \cr
\rho_{\rm repu.}(p) = & {1 \over {2\pi}} + O(n/c) . \cr}}
This implies that the free energies in this limit are
\eqn\eFreeBA{\eqalign{
f_{\rm attr.} ={E_{\rm attr.}\over L} - \mu n =  & {1 \over 6}\pi^2 n^3 - (\mu
+ 2c^2) n ,\cr
f_{\rm repu.} ={E_{\rm repu.}\over L} - \mu n = & {8 \over 3} \pi^2 n^3 - \mu n
.\cr
}}
These solutions have a simple physical interpretation.  On the
repulsive side, as $c \rightarrow \infty$, the two types of steps
are mutually avoiding, and may be mapped
onto a {\it single} species of Fermion, but with twice the density.
This is reflected in the equivalence of $\rho$ to its free value, and
to the consequent doubling of the Fermi momentum.  On the attractive
side, the up and down steps become tightly bound, and these bound
states can be treated again as free Fermions as $c \rightarrow
-\infty$.  In this case there are half as many total particles, but
only one species with twice the density, as reflected in the value of
$\rho$. The additional $-2c^2n$ in the free energy represents the
binding energy of these pairs.  In both cases, as the density goes
through zero, the conventional CIT
results, $n \sim \mu^{1/2}$, and $f \sim \mu^{3/2}$, are regained.

In the limit of weak coupling, the integral eqs.\eInteqnrep\ and
\eInteqnatt\ are extremely singular, rendering perturbative techniques
difficult.  We numerically solved the Bethe ansatz equations to see
the critical behavior as $c \goes 0$.  A plot of the ``heat capacity''
($\del^2 f / \del c^2$) is shown in Fig.[6].  No non--analytic behavior is
noticeable from the numerical solution despite the apparently
singular nature of the equations in this limit.  In fact, there may
be essential singularities which are impossible to discern from a
numerical solution (or experimental measurement).  This is the case
for the Kosterlitz--Thouless transition where the free energy has
singularities of the form $f
\sim \exp( -B/\sqrt{T - T_c} )$.\rKT

\newsec{Fermion Renormalization Group}

Another problem with the Bethe ansatz solution is that correlation
functions are not directly calculable.\foot{\dagger}{There has been a great
deal of effort devoted to this problem in the literature. We will
discuss later an indirect approach, but readers interested in exact
calculations should see, for example ref.[\the\refno]
\nref\rCorrfnctns{See, for example, V. E. Korepin, Comm. Math. Phys.
{\bf 113}, 177 (1987); and references therein.}.} We will
circumvent these difficulties through the use of a renormalization
group (RG) calculation around $c=0$.

There has been a great deal of interest recently in the extension of
RG methods to Fermion systems in more than one
dimension.\ref\rShankar{R. Shankar, Physica A {\bf 177}, 530 (1991);
and references therein.}\ In one dimension, such RG methods have a
longer history.  For this problem, Solyom\ref\rSolyom{J.  Soly\'om,
Adv. in Phys. {\bf 28}, 201 (1979).}\ has used both a `poor man's
scaling' and a field--theoretic correlation function type approach to
derive recursion relations.  For the sake of the statistical mechanics
audience, and for the relation to currently interesting extensions in
higher dimensional systems, we present an independent derivation using
the momentum--shell functional integral approach used by
Shankar.\rShankar

In the standard functional integral formulation of quantum field
theory, ground state expectation values (correlation functions) can be
expressed in terms of functional integrals with a weight given by the
action for the classical field theory.  In the statistical
mechanical problem of surface roughening, it is indeed only these
ground state expectation values which contribute in the large system
size limit.  For Fermionic field theories, it is necessary to use the
Grassman functional integral.\ref\rGrassman{See, for instance, L.
Ryder, {\it Quantum Field Theory,} (Cambridge Univ. Press, Cambridge,
1987), p. 187.}\ In this case, the partition function is given by
\eqn\eFint{ Z = \int\prod_{\alpha}
[d\psi^{\dagger}_{\alpha}][d\psi_{\alpha}] \, \exp(-S) ,}
\eqn\eAction{ S = \int \!\! dx dz \, \psi_{\alpha}^{\dagger}\!\left({
\partial_{z} - \partial^{2}_{x} - \mu} \right) \! \psi_{\alpha} + 2c
\int \!\! dx dz \, \psi^{\dagger}_{\uparrow}  \psi^{\dagger}_{\downarrow}
\psi_{\downarrow} \psi_{\uparrow} ,}
where it should be remembered that $\psi^{\dagger}$ and $\psi$ are
{\it independent} Grassman variables, and greek indices are summed
unless otherwise indicated.  Because of the anticommuting nature of
the Grassman fields, the definition of Fourier transforms requires
some care. We choose the following conventions:
\eqn\eFoura{ \psi (x,z) = \int \! {{dp d\omega} \over (2\pi)^2} \,\psi
(p,\omega) \exp
( ipx \! + \! i\omega z) ,}
\eqn\eFourb{ \psi^{\dagger} (x,z) = \int \! {{dp d\omega} \over
(2\pi)^2}  \,\psi^{\dagger} (p,\omega) \exp
( -ipx\! - \! i\omega z) .}
With these conventions the action becomes
\eqn\eActionma{ S = \int \! {{dp d\omega} \over (2\pi)^2} \,(i\omega + p^2 -
\mu)\,\psi^{\dagger}_{\alpha} (p,\omega) \psi_{\alpha} (p,\omega) \; +
\; V ,}
where
\eqn\eActionmb{ V = 2c\int \! {{d^{4}\!p d^{4}\!\omega} \over (2\pi)^8}
(2\pi)^{2} \delta ( \! p_1 \! + \! p_2 \! - \! p_3 \! - \! p_4 \!)
\delta (\! \omega_1 \! + \!
\omega_2 \! - \! \omega_3 \! - \! \omega_4 \!)
\psi^{\dagger}_{\uparrow}(p_1,\omega_1\!)
\psi^{\dagger}_{\downarrow}(p_2,\omega_2\!) \psi_{\downarrow}(p_3,\omega_3\!)
 \psi_{\uparrow} (p_4, \omega_4\!) .}

In conventional momentum shell RG calculations, the
primary concern is with the behavior at zero momentum, and therefore
the action is expanded around this point.  For Fermionic systems,
however, the long wavelength properties are determined by the behavior
near the Fermi momentum.  Intuitively, a change in the parameters of
the action below the Fermi surface has no effect on the low energy
properties of the system, since these levels are always completely
full.  There are thus two effectively independent fields near the two
Fermi points in one dimension.  To be consistent, some form of
momentum cut--off must be used both below and above the Fermi
points, so  as not introduce too many degrees of freedom or extra
couplings into the problem.  This approximation is illustrated in
Fig.[7].  To analyze more easily the modes near $\kf$, we linearize
the spectrum around these points.  The transformation to the new
fields is expressed as
\eqn\eTransf{\eqalign{
\psi(\pm p) = & \, \, \eta_{\pm} (\pm p - \kf ), \cr
\psi^{\dagger}(\pm p) = & \, \, \eta_{\pm}^{\dagger} (\pm p - \kf ) ,\cr
}}
where $p > 0$.  The linearized quadratic part of the
action is given by
\eqn\eQuadlin{S_0 = \int_{|p| < \Lambda} \meas \!\!\left[ (i\omega + \! 2\kf
p - \!m) \eta^{\dagger}_{+\alpha}\eta_{+\alpha}  +
(i\omega - \!2\kf
p - \! m) \eta^{\dagger}_{-\alpha} \eta_{-\alpha}
\right] ,}
where $\alpha = \uparrow,\downarrow$ is summed over the two Fermion
species.  Here a constant term representing the unperturbed energy of
the Fermi sea has been subtracted, and must be included in the free
energy.  The term proportional to $m$ reflects the freedom to choose
the point of linearization.  $m_{0}$, the initial value of $m$, must
be chosen order by order so that the renormalized mass goes to zero.
The necessity to choose a non--zero $m_{0}$ is simply a reflection of
the fact that the interactions change the relationship between the
Fermi energy and the Fermi momentum, the one--dimensional analogue of a
shift in the Fermi surface.

The interaction term is too cumbersome to write out algebraically.
It is instructive to consider a general model with more than the
original delta--function (Hubbard type) potential.  In this case there
are four possible couplings at quartic order, since the interaction
can involve both left and right moving particles.  There is a
standard notation for these interactions in the literature,
known as `g--ology', because the various couplings are
denoted by a set of $g_{i}$.  Because we prefer to use coupling
constants that have a more direct diagrammatic meaning, we will not
use this convention, and instead denote our coupling constants by $c_{i}$.
Table [1] gives the correspondence between our couplings and those of
the g--ology school.  The definition of the $c_{i}$ is given
diagrammatically in Fig.[8].  For example, the first diagram implies
an interaction of the form
\eqn\eDiagram{ S' = c_1 \int\! d\!xd\!z \, \left[
\eta_{+\uparrow}^{\dagger}\eta_{-\downarrow}^{\dagger}
\eta_{-\downarrow}\eta_{+\uparrow} + (\uparrow \leftrightarrow
\downarrow) \right]. }

To perform the RG transformation,
modes with momenta lying in a shell of width $\Lambda l$ near $|p| =
\Lambda$ are integrated out and momenta and fields are rescaled
in such a way as to keep the linear terms in $p$ and $\omega$ fixed.
To lowest order the possibilities of field renormalization and of
nontrivial (imaginary time) dynamic scaling can be ignored.  The
rescalings are then given by
\eqn\eRescale{\eqalign{
p = & \,\, p' / b, \cr
\omega = &\,\, \omega ' / b, \cr
\eta = & \,\,\eta ' b^{3/2}, \cr
\eta^{\dagger} = &\,\, \eta^{\prime \dagger} b^{3/2} .\cr }}
Here $b$ is the momentum rescaling factor and is set to $b = e^l $.
With these rescalings, we must examine the relevance of the possible
terms in the action.  A simple analysis shows that the quadratic terms
are marginal, and that all higher order terms are irrelevant near the
free Fermion fixed point.\foot{\dagger}{Indeed, this remains true for
Fermions in all higher dimensions.\rShankar}  To find
out their behavior under the RG, we must proceed to one--loop order.
Details of the calculation are given in Appendix A.

The RG equations can be put into a more familiar form by defining
dimensionless couplings $x_i
\equiv c_i/(4\pi\kf)$ and the linear combinations $y
\equiv x_1 - x_4$ and $\tilde{y} \equiv x_1 + x_4$.  The recursion relations
then become
\eqn\eRecursionb{\eqalign{
\dot{y} = & \, \, -2x_{2}^2 ,\cr
\dot{x}_2 = & \, \, -2x_{2}y ,\cr
\dot{x}_3 = & \, \, 0 ,\cr
\dot{\tilde{y}} = & \, \,  0 .\cr}}
The first two are exactly the recursion relations for the Kosterlitz--Thouless
(KT) transition of the XY model in two dimensions! The role of the vortex
fugacity is played by $x_2$, and the reduced temperature corresponds
to $y$ (see Fig.[8]). The two terms $x_3$ and $\tilde{y}$ are marginal to
this order.  The fact that $x_2 = 0$ corresponds to the fixed line in
the XY model can be easily understood: in the absence of $x_2$
the theory has an additional symmetry corresponding to independent
conservation of spins at each Fermi point and this symmetry
is preserved under the RG.  This term corresponding to $x_2$ in fact
describes back--scattering in the Fermion terminology.

These recursion relations {\it do not} imply that the Hubbard model
has a Kosterlitz--Thouless transition as $c$ goes through 0.  The
initial conditions for the Hubbard model correspond to being on the
special line $y = x_2$ in eqs.\eRecursionb\ (see Fig.[9]).  In
fact, this equality is a consequence of the $SU(2)$ spin--rotation
invariance of the Hubbard model, as can be made manifest
by rewriting the interaction term as
\eqn\eIntsym{V = c\psi^{\dagger}_{\uparrow}  \psi^{\dagger}_{\downarrow}
\psi_{\downarrow} \psi_{\uparrow} = {c \over 2}
\left( \nup + \ndown \right)^{2}.}
To obtain the singularities within the Hubbard
model, note that the RG in this case reduces to a single
flow equation for one marginal parameter $x \equiv x_2 = y$,
\eqn\eFlowone{ \dot{x} = -2x^2.}
The singular part of the free energy obeys the scaling relation
\eqn\eFreescaling{ f_s(x_0) = e^{-2l} f_s\big(x(l)\big),}
where $x_0$ is the initial value of $x$.  For repulsive interactions,
$x(l) \goes 0$ as $l \goes \infty$, so that $f_s$ vanishes.  For
attractive interactions, $x$ is marginally relevant.  Choosing
$l=l^*$ such that $x(l^*) = 1$ and using the solution for
eq.\eFlowone, eq.\eFreescaling\ becomes
\eqn\eGap{ f_s(x_0) =  f_s(1)\exp( -1 - 1/x_0) = C\exp( -1/x_0).}
The relevance of $x$ on the attractive side simply signals the
formation of a bound state (mass gap), and could have been predicted
on general grounds.

The other two parameters $x_3$ and $\tilde{y}$ remain marginal at
second order, a result expected to hold to all
orders.\ref\rPRL{C. Di Castro and W. Metzner, Phys. Rev. Lett. {\bf
67}, 3852 (1991). }\ For repulsive couplings we flow to the fixed
sub--space where $x_2 = 0$, and we must solve this more general
problem, known as a Luttinger model.\ref\rLuttMatt{ Luttinger posed
this problem in J. Math. Phys. {\bf 4}, 1154 (1963).
The correct solution was found by Mattis and Lieb in ref.[28].}

\newsec{Luttinger Liquid and Correlation Functions}

This problem, in the linearized form used in the previous section, was
first solved for spinless particles by Mattis and
Lieb.\ref\rMattlieb{D.  Mattis and E. Lieb, J. Math.  Phys. {\bf 6},
304 (1963).}\ They found that any such model with $x_2 = 0$ can be
solved exactly in terms of Boson operators, essentially because the
interactions can be written entirely in terms of density operators for
the different Fermion species.  Mattis and Lieb showed that these
density operators, properly normal--ordered, have exactly the
commutation relations of Boson creation and annihilation operators,
and that the free--Fermion Hamiltonian can be written as a quadratic
form in these operators.  The interacting model can then be easily
solved by a Bogoliubov transformation.  The singular behavior at $p =
0$ is unchanged, while the divergences at the Fermi momentum change
continuously with the interaction strength.

Mattis and Lieb worked out this equivalence at the level of creation
and annihilation operators.  In the second quantized notation, the
formulae simplify a great deal.  The Fermion Hamiltonian for the
spinless case is a simplified version of eq.\eQuadlin.\foot{\dagger}{In the
context of field theory, this is the Hamiltonian of a single Dirac
Fermion.}  In real space,
this is
\eqn\eMajorana{ H = iv \int \! dx \left( \eta^{\dagger}_{+}\del_{x}\eta_{+} -
\eta^{\dagger}_{-}\del_{x}\eta_{-} \right) .}
The statement of Bosonization is that the
Hamiltonian for this model can be written in terms of a scalar field
as
\eqn\eLuttliquid{H = {v \over 2} \int dx \left[ {\pi K} \Pi^2 + {1
\over {\pi K}}
(\del_{x}\phi)^2 \right],}
where $\Pi = \del_{z}\phi/(\pi Kv)$ is the canonical momentum.  For
the free Fermion Luttinger model, $K=1$ and $v = 2\kf$. The
Fermionic field operators $\eta_\pm$ have nonlocal expressions in terms of
$\phi$:
\eqn\eBosonequiv{ \eta_{\pm} (x) \propto \,\, \,\exp\left( \,\pm i\phi (x) -
i\pi\int_{-\infty}^{x}\!\! dy \Pi(y) \right) . }
Using eq.\eBosonequiv, interactions added to the Fermion Hamiltonian
\eMajorana\ can be written in the Bosonic language.  Some terms
(e.g. $\eta_{+}^{\dagger}\eta_{+}\eta_{-}^{\dagger}\eta_{-}$) can be
absorbed into the quadratic Boson Hamiltonian, indicating that the interacting
Fermion system is {\it exactly}
equivalent to free Bosons, as implied by the
solution of Mattis and Lieb.  For other types of interactions, the
corresponding Bosonic terms {\it are} higher order than quadratic.
Even in this case, the interactions may be irrelevant in the
RG sense, so that the system flows to the free
field fixed point.  In either case, the only difference between the
different Fermion theories in the Bosonic language is that the
parameters $K$ and $v$ do not take on their free Fermion values.

The method is easily extended to the problem of spin--1/2 Fermions
by using two Bosonic fields, $\phi_{\uparrow}$ and $\phi_{\downarrow}$.  This
provides, for one, an alternative method for performing the
RG calculations of the previous
section.\foot{\dagger\dagger}{See appendix B for a sketch of this
calculation.} Additionally, on the fixed line $x_{2} = 0$, where a
Bogoliubov transformation solves the Luttinger model, the Bosonized
theory is already free!  The only complication is that the parameters
$K$ and $v$ take on values different from those corresponding to
free Fermions.  This is the essence of the ``Luttinger liquid''
theory of Haldane\rHaldane --- that spin--1/2 Fermions (and
indeed a large class of models) are within the basin of attraction of
the fixed surface described by a pair of free scalar fields.  For
sufficiently low energies, the original model renormalizes
down to the Luttinger liquid fixed line and its correlation functions
are calculated  by computing Bosonic expectation values.
On this fixed subspace, the theory takes a simple Gaussian form
\eqn\eLuttliquidtwo{H = \sum_{\nu = \rho,\sigma} {1 \over {2\pi
K_\nu}} \int dx
\left[ {1 \over v_\nu} (\del_z \phi_\nu)^2 + v_\nu
(\del_{x}\phi_\nu)^2 \right], } where $\rho$ and $\sigma$ refer to
charge and spin fields respectively, with corresponding parameters
$K_{\rho},v_{\rho}$ and $K_{\sigma},v_{\sigma}$. In the fully
renormalized theory of eq.\eLuttliquidtwo, the only effect of the
interactions of the original model is to change the values of the
parameters $K_\nu$ and $v_\nu$.
The spin--spin correlation function has the asymptotic
form\ref\rSchultz{T. Giamarchi and H.J. Schulz, Phys. Rev. B {\bf
39}, 4620 (1989).}
\eqn\eSscorr{\langle S(x)S(0) \rangle \sim {K_\sigma \over (\pi
x)^{2}} - A{{\cos (2\kf x)} \over x^{K_{\sigma} + K_{\rho}}},} for
large $x$, where $A$ is a non--universal coefficient.  For the $SU(2)$
symmetric case $K_\sigma = 1$,\foot{\dagger}{Logarithmic corrections
are present in this case from the marginally irrelevant operator.\rSchultz}
but it can take on any value between
zero and one in general. For the crystal surface, $S(x)$ is the slope
at the position $x$, and using eq.\eHeightdiff\
the physical correlation function is
\eqn\eHeightcorr{ C(x,0) = \left\langle \left(h(x,z) - h(0,z)\right)^2
\right\rangle =
\int_{0}^{x} \!\!dy dy' \,\langle S(y,z)S(y',z) \rangle .}
By changing to center of mass variables, this becomes
\eqn\eHeightcom{ C(x,0) = 2\int_0^x \!\!dz \, (x-z)\langle
S(y,z)S(y',z) \rangle.}
Differentiating twice yields a linear equation for $C(x,0)$, whose
asymptotic solution has the form
\eqn\eHeightlog{C(x,0) \sim {2 K_\sigma \over \pi^2}\log x + {A \over
{2\kf^2}} {\cos(2\kf x) \over {x^{K_\sigma + K_\rho}}}.}
The reconstruction order parameter correlations are computed in a similar
manner and behave as
\eqn\eRoughcorrlead{ G(x,0) \sim \left( {1 \over x}
\right)^{K_\rho/p^2} \cos\left( {{2\pi n x} \over p} \right). }

In the flat incommensurate phase (i.e. for sufficiently attractive
potentials), the theory is no longer within the domain of attraction
of the spin--$1/2$ Luttinger liquid fixed point in eq.\eLuttliquidtwo .
Nevertheless, the correlations are described by a Luttinger liquid
type model.  In the limit of large attractive
interactions, the system is well described by an assembly of tightly
bound up--down pairs.  Thus a {\it spinless}
Luttinger liquid description is valid for these bound states, as is
generically the case for one--dimensional quantum systems.\rHaldane\ By
continuity, such a description is valid at low energies throughout the
entire smooth phase.

\newsec{ Renormalization Group for Generic Interactions }

The choice of a delta--function interaction between lines results in
the $SU(2)$ symmetry of the Hubbard model, as shown in section IV.
This additional symmetry restricts the RG flows to
lie precisely along the critical line of the Kosterlitz--Thouless phase
diagram.  Since this line is never crossed, the KT--type singularities
are never observed.  The most general form of a translationally invariant
two--body interaction with this symmetry is constructed using the
singlet density operator $n_{\rho}
\equiv \nup + \ndown$ as
\eqn\eHintsing{H_{\tiny SU(2)} = \int \! dx dy
\, n_{\rho}(x) V(x-y)n_{\rho}(y). }
Since in general, step--step interactions do not have this $SU(2)$
symmetry, RG again determines the generic
behavior.  Eqs.\eRecursionb, still valid in this case,
indicate a wedge of stability for the fixed line, depicted in Fig.[9].
Since this wedge is of the same dimension as the full parameter space,
this phase is expected to be present in a generic system. For example,
consider a finite range repulsive force between up and down steps
described by a potential
\eqn\eExplicitpot{ V_{\uparrow\downarrow}(x) = \exp\big(- {1 \over
2}(x/a)^2 \big).} At least within the perturbative RG, this potential
is within the wedge of stability for $a>0$ and on the separatrix in
the limit of $a=0$.  For any such system not satisfying the $SU(2)$
symmetry, a transition between the bound and unbound phases exhibits
normal KT behavior. This is because the trajectory representing the
potential will most likely cross the separatrix, rather than take a
pathological path through the origin of the phase plane.  Even for
these systems, the Luttinger liquid treatment still holds,
since the final renormalized theory falls on the fixed line.

The RG treatment is valid for long--range interactions as well, with
certain restrictions.  In general, it holds for sufficiently weak
potentials if the four point interactions have no singularities in
momentum space near $\pm\kf$, and if they remain bounded so that it is
appropriate to perturb around a Fermi sea.  The first potential
problem corresponds to potentials which oscillate with a period
commensurate with the inter--step distance.  This is probably unlikely
for any finite density of steps, and will be ignored here.  The second
possibility occurs for sufficiently long--ranged potentials, such
as an isotropic interaction decaying as
\eqn\eLongrangex{ V(x,z) \sim {1 \over (x^2 + z^2)^{\gamma/2}} \equiv
{1 \over r^\gamma}.}
The Fourier transform of this potential has the small momentum
scaling
\eqn\eLongrangep{ \tilde{V}(p) \sim p^{\gamma - 2},}
which remains finite at small $p$ for $\gamma > 2$.
The RG treatment should thus hold for all
potentials decaying faster than $1/r^2$.  Typical elastic surface
potentials decay like $1/r^3$, so our treatment is
justified.\foot{\dagger}{The potential between {\it units} of the
steps scales like $1/r^3$, so that the full interstep potential scales
like $1/r^2$. }


\newsec{ Introduction of Loops }

For a $\pbyo$ reconstructed surface, it was noted in section I that it
is possible for the $2p$ steps to intersect forming a ``loop''.  For
large $p$, the probability of $2p$ lines intersecting near a single
point becomes very small, so that such configurations are irrelevant.
We will show explicitly that this is true for $p>2$, in which case the
previous treatment is applicable, while for $p = 1,2$ more care is
necessary.  To introduce loops into the model, the operators that
create the bottoms and tops of the loops (see Fig.[10]) must be added
to the action.  The proper choice of such operators, as noted by
Halperin {\it et. al.} in the case of the CIT,\rHalperin is a product
of $2p$ Fermion creation or annihilation fields, i.e.
\eqn\eSloop{S_{\tiny LOOP} = \lambda\int \! dx dz \, \left[
\prod_{k=0}^{p-1}\psi_{\uparrow}(\!x\!+\!ka\!)\psi_{\downarrow}(\!x\!+\!ka\!)
 + h.c.
\right] ,}
where $a$ is a small distance introduced to keep the Fermion operators
(all at equal $z$) separate.  Upon expansion, this action represents a
grand canonical description in the loop fugacity $\lambda$.

Power counting immediately shows that $\lambda$ is irrelevant for
$p>2$, since it accompanies more than four Fermion operators.  The
phase diagram in the absence of loops should thus hold for such
systems.  This is in agreement with the experiments on the $3 \times
1$ gold (110) surface by Held et. al.,\rBirgeneau whose observations are
consistent with our predictions for the FO--RI transition.

For
$p=1$, loops are strongly relevant, and for $p=2$, $\lambda$ is again a
marginal four--Fermion interaction.  Re--expressing this interaction in
terms of the operators $\eta_{\uparrow}$ and $\eta_{\downarrow}$
results in only one term at zero momentum (i.e. with all four of the
original momenta at $\pm \kf$), shown in Fig.[11].  It is a simple
matter to include such a term in the RG equations, either by the
methods of section IV or by Bosonization techniques.  The new
interaction results in a second, decoupled set of KT recursion
relations
\eqn\eLooprecursions{ \eqalign{
\dot{\tilde{y}} = & \, \, -2w^2, \cr
\dot{w} = & \, \, -2w \tilde{y} ,\cr}}
where the dimensionless fugacity is given by $w \equiv
\lambda/(4\pi\kf)$.  Recall that for the Hubbard model, the initial
value of $\tilde{y}$ is simply proportional to $c$.  Thus for $c>0$,
the system is stable to loops for a finite range of fugacity (to
lowest order this is given by $|w|<\tilde{y}$).  For larger fugacities
loops become relevant even for $c>0$, indicating a transition in the
KT universality class.  For $c \leq 0$, loops are always relevant, and
it is easy to understand the instability: in this regime, the Fermion
lines are paired up into bound states of up and down spins.  As these
bound states act effectively as single lines, the requirement of four
Fermion lines meeting at a vertex reduces to only two bound pairs
intersecting.  This is a much more probable event; indeed, from the
above power counting argument (or from knowledge of the
CIT\rHalperin), such operators {\it are} relevant.  The stability for
large positive $c$ is also easily understood, as in this case the
lines are completely mutually avoiding. Again, by analogy to the CIT,
this system is stable to vortices for $2p>2\sqrt{2}$.  The relationship
between the original derivation of this result by Coppersmith {\it et.
al.}\rCoppersmith\ and our approach is described in appendix C.

Since the smooth phase is always unstable to loops for $p=2$, the
critical behavior of the FO--FD transition is different from the CIT
behavior of the FO--FI transition it replaces.  Indeed, in the limit of
large negative $c$, the system is composed of tightly bound pairs of
up and down steps, which may be treated as weakly interacting
Fermions.  The loops in this language come from pairs of
creation or annihilation operators.  This is precisely the transfer matrix
description of the two dimensional Ising model.\ref\rIsingtwod{T.
Bohr, Phys. Rev. B {\bf 25}, 6981 (1982); M. Kardar and R. Shankar,
Phys. Rev. B {\bf 31}, 1525 (1985).}\  Thus this
transition is Ising--like, as also expected on
symmetry grounds once spatial isotropy is regained by the introduction
of loops.

For the $p=2$ case of gold (110), there are two possible scenarios as
a function of {\it temperature}, corresponding to the two possible
trajectories in parameter space shown in Fig.[12].  In the
first case (trajectory (a)), the surface undergoes an Ising
deconstruction transition to a flat disordered state, followed by an
ordinary roughening transition at higher temperatures.  This is
apparently the case for gold (110).\rGoldexpb\ The other situation,
which, although not realized in gold (110), may be present in other
systems, follows from path (b) in Fig.[12].  The surface first
undergoes a CIT to the rough incommensurate phase, followed by a
KT$^*$ transition in which the {\it reconstruction} order parameter
has a universal jump in the associated stiffness $K_\rho$.  Since the
spin--charge separation of section V is only a property of the low
energy Luttinger liquid theory, this KT$^*$ transition for the
reconstruction correlations may drive a simultaneous transition for
the height sector.  Whether this occurs cannot be determined from the
RG analysis, since it depends upon the strong coupling limit of the RG
flows.  If it does occur, the system goes directly to the flat
disordered phase, with a discontinuous jump in the height
correlations.  If this is not the case, there is an intermediate phase
in which the surface height remains rough but the reconstruction
correlations are disordered (RD).

Generically, this RD phase shares a boundary with the simple FO phase.
Unfortunately, this low density regime is outside the range of
applicability of the Fermionic RG.  Here, because the system is
isotropic, the results of den Nijs\rDenNijs\ for combined roughening
and deconstruction should apply.  From a numerical finite size study, he
concludes that such a transition exhibits superimposed roughening and Ising
critical behavior, with a central charge of $c=3/2$.  This situation
agrees well with experiments on Pt (110),\rRVK\ with the trajectory shown in
Fig.[2b].

\newsec{Conclusions}

We have studied the possible behaviors for roughening of
anisotropically reconstructed crystal surfaces.  For a general $\pbyo$
reconstruction, the phase diagram is strongly dependent on $p$.  For
$p>2$, loops are at least perturbatively irrelevant, and there are
three possible phases: flat ordered (FO), flat incommensurate (FI),
and rough incommensurate (RI).  The two transitions to the FO phase are in
CIT universality class, while the FI--RI transition is generically in
the KT class (a different essential singularity of the form
$\exp(-2\pi\kf / c)$ is found for systems with $SU(2)$ symmetry).
This situation is depicted in Fig.[5b].  For $p=2$, the phase diagram
is modified.  Four phases are now possible: flat ordered (FO),
flat disordered (FD), rough incommensurate (RI), and rough disordered (RD).
The FO--RI transition remains in the CIT class, the FO--FD transition is
in the Ising universality class, and the FD--RD and RI--RD transitions
are in the KT and KT$^*$ classes respectively, when the RD phase
exists. When it does not, the FD--RI transition will be of the KT$^*$
class, with a {\it non--universal} jump in $K_\sigma$ for the
height--height correlations.  This situation is depicted in Fig.[5c].
For $p=1$, the loops are always relevant, and the model has only a
single reconstruction transition (Fig.[5a]).  The real system in this
case may also have an ordinary roughening transition, though this is
not built into the model.

The experimental observations on gold (110) are consistent with one
theoretical scenario, but leave open the possibility of a different
phase diagram in other $2 \times 1$ reconstructed systems.  The
alternative situation predicts a new rough phase with logarithmic
height fluctuations and unusual singularities at non--zero momentum, as
well as floating incommensurate order in the reconstruction order
parameter.

Experimental observations of the Fermi surface singularities by
electron diffraction will probably be difficult, since the structure
function is essentially dominated by the small momentum behavior.  A
more promising route is scanning tunneling microscopy (STM),
by which direct measurements can be made of surface height
profiles.\rVicinal\ Such data can be directly Fourier transformed to
momentum space in order to search for Luttinger--liquid type behavior.
Since our treatment can also be carried through for the simpler case
of a vicinal surface, these singularities should also be observable
(and the observations will probably be easier) in such systems.  A
sketch of the Bosonization treatment for a vicinal surface is given in
appendix D.

There are probably many other applications of these methods to surface
physics.  It would be interesting to try and address the issue of
terrace--width distributions using these techniques.\rVicinal\ Other
possible applications involve step--doubling transitions,\ref\rSdoub{T.
L. Einstein, T.M. Jung, N.C. Bartelt, and E.D. Williams, to be
published in J. Vac.
Sci. Tech. A {\bf 10}, (1992). }\ and disordered systems.\ref\rUs{
L. Balents and M. Kardar, preprint, (1992).}\ We have also found it
interesting to understand better the connection between the various
treatments of commensurate to incommensurate transitions, roughening,
and interacting Fermions.

\centerline {\bf Acknowledgements}
We have benefited from discussions with X. Wen, O. Narayan, and R.
Kamien. We are very much indebted to J. Villain for both introducing us to this
problem, and for important comments on an early version of this manuscript.
This research was supported by the NSF through a Graduate
Fellowship (LB), and by grant number DMR--90--01519 and the PYI program
(MK).

\appendix{A}{One Loop Renormalization Group Calculation }

A priori, this appears to be a complex calculation, as the four
different vertices allow for $ {4 \choose 2} + 4 = 10$ different
pairings, each of which can result in several distinct diagrams
depending upon the choice of contractions.  A great simplification is
achieved by examining the possible loops that may be formed from two
such vertices.  Since field renormalization does not occur at this
level, the diagrams need only be computed in the zero momentum limit,
which means that the momenta on each half of a loop must be equal. In
the field--theoretic language, the required diagrams are known as
four--point vertex functions at zero momentum.  A
simple calculation then shows that of all the possible (6) loops that
may be drawn, only two are non--zero.  All the others vanish because
both poles are in the same half--plane.\foot{\dagger}{The sign of the
momentum is important, since it appears linearly.  It is determined by
checking whether the momentum is directed along or against the arrow
of the propagator line.} Physically, this is a reflection of
causality.  The linear propagators represent ballistic propagation in
real space.  The two non--zero loops give
\eqn\eLoops{|\Gamma^{4}_{+-}(0)| = {l \over {2\pi\kf}} +
O(m/(\kf\Lambda)),}
where $\Gamma^{4}_{+-}(0)$ is the four--point vertex function at zero
momentum for one left-- and one right--moving Fermion.  The diagram with
the arrows parallel is negative, while the antiparallel diagram is
positive.

With these simplifications, the computation of the one--loop diagrams
is quite straightforward.  The only remaining complexity is in
handling the Fermi minus signs.  Curiously, there are {\it no} minus
signs at the one--loop level.  This follows from the fact that, in
momentum shell RG, the uncontracted fields are always left on the
diagrams, since they must be put back into the exponential.  One can
easily convince oneself of this by calculating a few of the terms.
The final result is quite simple
\eqn\eRecursiona{\eqalign{
\dot{x}_1 = & \, \, -x_{2}^2, \cr
\dot{x}_2 = & \, \, -2x_{2}(x_1 - x_4), \cr
\dot{x}_3 = & \, \, 0, \cr
\dot{x}_4 = & \, \,  x_{2}^2,\cr }}
where $c_i = 2\pi\kf x_i$ defines the dimensionless couplings $x_i$.

\appendix{B}{Renormalization Group Equations Via Bosonization }

In order to perform the RG calculation in the
Bosonic Hamiltonian, it is first necessary to absorb the vertices of
the original model which do not correspond to interactions in the
Bosonic language.  This is best done while still in the Fermionic
variables, so that one does not need to deal with divergences relating
to the coefficient of the transformation \eBosonequiv.  The
parameters of the Bosonized Hamiltonian are related to those of the
Fermionic one via
\eqn\eParameters{\eqalign{
K_{\rho} = & \, \, \left( {1 - \tilde{y}} \over {1 + \tilde{y}}
\right)^{1/2}, \cr
v_{\rho} = & \, \, \kf\left( (1 + x_3)^2 - \tilde{y}^2 \right)^{1/2}, \cr
K_{\sigma} = & \, \, \left( {1 + y} \over {1 - y}
\right)^{1/2}, \cr
v_{\sigma} = & \, \, \kf\left( (1 - x_3)^2 - y^2 \right)^{1/2}. \cr }}

Once these density--density interactions have been accounted for, only
the vertex corresponding to $x_2$ remains.  It cannot be written in a
form quadratic in the field operators, and thus must be treated as an
interaction.  Using eq.\eBosonequiv, the corresponding interaction in
the Bosonic language is easily found to be
\eqn\eBosonxtwo{ H_{\rm int} = x_2 \int \! dx \, \cos
(2\sqrt{2}\phi_{\sigma}), }
where $\lambda \propto x_2$, and  charge and spin Bosons are defined by
\eqn\eSpincharge{\eqalign{
\phi_\rho = & \, (\phi_{\uparrow} + \phi_{\downarrow})/\sqrt{2}, \cr
\phi_\sigma = & \, (\phi_{\uparrow} - \phi_{\downarrow})/\sqrt{2}.
\cr}}
The charge Boson is now completely free and decoupled from the spin
sector, while the spin Hamiltonian contains the interaction
\eBosonxtwo .  In fact, the resulting Hamiltonian is precisely that of
the sine--Gordon model.\ref\rSinegordon{S. Coleman, Phys. Rev. D {\bf
11}, 2088 (1975).}\  This theory is well known to be dual to the
XY model with vortices,\ref\rJose{ J.V. Jos\'e, L.P. Kadanoff,
S. Kirkpatrick, and D.R. Nelson, Phys. Rev. B {\bf 16},
1217 (1977).}\ with the KT critical point located at $K = 1$
and $\lambda = 0$.  The RG flows for small $x_2$
are identical to those derived within section IV in the Fermion
theory, thus providing an independent derivation.

It is not a coincidence that the sine--Gordon theory arises also in the
conventional theory of the roughening transition.  We can see the
correspondence more explicitly by constructing the height operator for
the crystalline surface.  The density of steps (Fermions) of species
$\alpha$ has the simple expression in terms of the Bosonic fields,
\eqn\eStepdensity{ n_\alpha (x) = n_0 + {1 \over \pi}\del_x
\phi_{\alpha} + M\cos(2\phi_\alpha + 2\kf x), }
where $n_0$ is the average density, and $M$ is a cut--off dependent
constant.  From this, the height operator can be constructed according
to eq.\eHeightdiff .  Using eq.\eStepdensity, and the definition of
the spin Boson in eq.\eSpincharge,
\eqn\eHeightBoson{ h(x,t) = {\sqrt{2} \over \pi}\phi_{\sigma} +
2M\int_{-\infty}^{x} \!\! dy \, \cos\big(\sqrt{2}\phi_\rho(y,t) + 2\kf
y\big)\cos\big(\sqrt{2}\phi_\sigma(y,t)\big). } The free Hamiltonian $H_0$
(neglecting the non--linear corrections from the cosine parts of
eq.\eHeightBoson) is just the usual surface tension term, while the
interaction $\cos (2\pi h)$ in eq.\eBosonxtwo\ represents the
discrete translational symmetry by the step height (which equals $1$
in this model).

\appendix{C}{Loop Stability Analysis by the Displacement Field Method }

Coppersmith {\it et. al.}\rCoppersmith\ originally performed the stability
analysis for the CIT via a method different from the one used in
section VII.  Their approach is to first find the effective
displacement field theory applicable to the CIT and introduce loops as
dislocations in the line lattice; i.e.,  vortices in the
(single--component) displacement field.  Before loops are introduced,
for $c \geq 0$, an effective displacement field theory can be
developed for the two species of steps.  At $c=0$, the effective free
energy has the form
\eqn\eEffectF{ F = \sum_{\alpha = \uparrow,\downarrow} \int \! d\!x
d\!z \, \left[ {B_x \over 2} (\del_x u_\alpha)^2 + {B_y \over 2}
(\del_z u_\alpha)^2 \right]. } Only the combination $B = \sqrt{B_x
B_y}$ is necessary for the stability analysis; it is fixed by matching
the displacement correlations in the effective theory to those
calculated directly in the Fermion language.  In the effective theory,
\eqn\eGeff{ G(x) \equiv \left\langle \bigg(u_\uparrow (x) - u_\uparrow
(0) \bigg)^2
\right\rangle = {1 \over {\pi B}} \log ({x\over a}), }
where $a$ is the short--distance cut--off.  In the free Fermion
language, the same quantity is given by
\eqn\eGferm{ G(x) = \left\langle \left( \int_{0}^{x} \! dy \, n_\uparrow
(y) - nx \right)^2 \right\rangle = {1 \over \pi^2} \log ({x\over a}).}
Thus we have $B = \pi$ for $c=0$.  The linear combinations of $\uup$
and $\udn$, such that the loop of $2p$ lines corresponds to a
``vortex'' in just one of these fields, are
\eqn\eLincomb{\eqalign{
\tr \equiv & {\pi \over p} (\uup + \udn), \cr
\ts \equiv & {\pi \over p}(\uup - \udn) . \cr}}
As the closed path through the loop in Fig.[10] is traversed, the two
new fields shift by $\tr \rightarrow \tr + 2\pi$ and $\ts \rightarrow
\ts$.  This is just a standard singly charged vortex in the $\tr$
field.  In terms of these fields, $F$ separates into ``charge'' and
``spin'' parts.  The ``charge'' part, which is important to our
analysis, is
\eqn\eFcharge{ F_\rho = {1 \over 2} \left( {p^2 \over {2\pi^2}}
\right) \int \! d\!x d\!z \, \left[ B_x (\del_x \tr)^2 + B_y (\del_z
\tr)^2 \right].}
The ``temperature'' of this XY model is $T_\rho = 2\pi^{2}/(p^{2}B) =
2\pi/p^2$ at $c=0$.  For $p=2$, this is precisely at the KT critical
point, so the system is marginally unstable to these types of loops.
In fact, spin--charge separation implies that this form will remain
true for $c>0$, but with a different value of $T_\rho$.  The KT
analysis states that the system is stable for $T_\rho < \pi/2$, so the
previous analysis implies that this condition must be satisfied for
$c>0$.  For $p>2$, it is satisfied even for $c=0$.

We can understand the result for $p=2$ a little more deeply by
considering the Bosonization treatment of the loops using
eq.\eBosonequiv.  In this case, the charge Boson $\phi_\rho$ is
described by a Hamiltonian
\eqn\eHrho{ H_\rho = {1 \over {2\pi K_\rho}} \int \! dx \, \left[
v_\rho (\del_x \phi_\rho)^2 + {1 \over v_\rho} (\del_z \phi_\rho)^2
\right] + \lambda \cos \left( {{2\sqrt{2}} \over {K_\rho v_\rho}}
\int^{x} \! d\!y \, \del_z \phi_\rho (y) \right). }
The change of variables
\eqn\eWchange{ \tilde{\phi}(x,z) \equiv {1 \over {K_\rho v_\rho}} \int^{x} \!
d\!y \,
\del_z \phi_\rho (y,z), }
results in a much simpler form for $H_\rho$.  Substitution of
eq.\eWchange\ into eq.\eHrho, along with use of the equation of motion for
$\phi_\rho$, yields the Hamiltonian in terms of $\tilde{\phi}$
\eqn\eHrhow{ H_\rho = {K_\rho \over {2\pi}} \int \! dx \, \left[
v_\rho (\del_x \tilde{\phi})^2 + {1 \over v_\rho} (\del_z \tilde{\phi})^2
\right] + \lambda \cos (2\sqrt{2} \tilde{\phi}) .}
This is once again the Hamiltonian of a sine--Gordon model.  Notice
that in this treatment, the loop interaction appears as a cosine term
rather than as a vortex operator.  In fact, these two models are known
to be related by a simple duality mapping!\rJose\
Under this mapping, the sine--Gordon theory of eq.\eHrhow\
becomes precisely the XY  model with vortices.  The corresponding
XY temperature is $T_\rho = (\pi/2) K_\rho$, so that the stability criterion
$T_\rho < \pi/2$, corresponds to the Luttinger liquid result
$K_\rho < 1$.

\appendix{D}{Bosonization for Vicinal Surfaces}

The correlation functions for vicinal surfaces can also be obtained
from the Fermion mapping.  In this case, there is only a single
species of Fermion, corresponding to the fact that a miscut surface
is composed of predominantly one type of step.  The low energy
action for such a theory is
\eqn\eVicinalaction{S = {1 \over {2\pi K}}\int \!\! d\!x d\!z \, \left[
v(\del_x\phi)^2 + {1 \over v}(\del_z\phi)^2 \right] ,}
where $v$ is the renormalized Fermi velocity, and $K$ is the Luttinger
liquid parameter.  $K>1$ for attractive interactions, $K<1$ for
repulsive interactions, and $K=1$ for free Fermions.  The deviation
from the average step density is given by
\eqn\eDeltan{ \delta n(x,z) \equiv n(x,z) - n_0 ={\del_x \phi(x,z)\over\pi}+
M\cos\big(2\phi(x,z) + 2\kf x\big),}
where $n_0$ is the average density of steps, and $M$ is a cut--off
dependent constant.  The density--density
correlations are then easily computed in the form
\eqn\eDensdensvic{\langle \delta n(x,z) \delta n(0,z) \rangle \sim -{K
\over {2\pi^2 x^2}} + \tilde{A} {\cos(2\kf x) \over x^{2K}},}
where $\tilde{A}$ is a constant proportional to $M$.  Using this
expression and the methods of section V, the height--height
correlations are
\eqn\eVicheight{C_v(x,0) \equiv \left\langle \big(h(x,z) - h(0,z) -
n_0 x\big)^2
\right\rangle \sim {K \over \pi^2}\log |x| - A{\cos(2\kf x) \over x^{2K}},}
where $A$ is another constant.  In momentum space, the Luttinger
liquid singularities are
\eqn\eVicheightmom{C_v(p) \sim C + D|p \mp 2\kf|^\alpha \,\,\, {\rm
as} \, p \goes \pm 2\kf.}
The parameter $\alpha>1$ for attractive interactions, $\alpha<1$ for
repulsive interactions, and $\alpha=1$ for free Fermions.  In contrast
with the roughening case, this exponent and the coefficient of the
logarithmic term in the roughness {\it are not independent}; in fact,
\eqn\eRelated{ \alpha = 2K -1 .}

\vfill\eject\immediate\closeout\rfile                       %
\centerline{{\bf References}}\bigskip                                     %
{\catcode`\@=11\escapechar=`  \input refs.tmp\vfill\eject}

\vfill\eject
\vfill

\eject
\vfill
\vbox{\magnification=1000\tabskip=0pt \offinterlineskip
\def\tablerule{\noalign{\hrule}}
\halign to 2.4in{\strut#& \vrule height20pt#\tabskip=2em minus2em&
\hfil#\hfil &\vrule# &\hfil#\hfil &\vrule#\tabskip=0pt\cr
\tablerule
&&Our Notation &&g-ology&\cr\tablerule
&&$c_1$&&$g_2$&\cr\tablerule
&&$c_2$&&$-g_{1\perp}$&\cr\tablerule
&&$c_3$&&$g_4$&\cr\tablerule
&&$c_4$&&$g_2 - g_{1\parallel}$&\cr\tablerule
}}
\vfill
\tab{1}{Relationship between our coupling constants and those of the
g--ology description.}
\vfill
\psfig{figure=fig1.ps}
\fig{1}{Geometry of the $2 \times 1$ reconstructed gold (110) surface.
}
\vfill
\psfig{figure=fig2.ps}
\fig{2}{Four possible sublattices on a $2 \times 1$
reconstructed surface.
In general there are $2p$ such sublattices for a $\pbyo$
geometry.}
\eject
\psfig{figure=fig3.ps}
\fig{3}{A pair of up and down steps on the $2 \times 1$ surface.
After such an up/down pair, the top row of atoms moves to the alternate
sublattice.}
\vfill
\psfig{figure=fig4.ps}
\fig{4}{Smallest allowed loops for (a) $p = 1$ , (b) $p = 2$, and (c) $p =
3$. }
\vfill\eject

\psfig{figure=fig5.ps}
\fig{5}{Phase diagrams for (a) $p = 1$, (b) $p > 2$, and (c) $p  = 2$.
The phase boundaries reflect only the topology and not the detailed
shape of the phase space.
}\vfill\eject

\psfig{figure=fig6.ps}
\fig{6}{``Heat Capacity'' ($-\del^2 f / \del c^2$) for
spin--$1/2$ Fermions
with a delta function potential calculated via the Bethe ansatz for a
denisty $n=1$. Eq.\eFreeBA\ implies an asymptotic value as $c
\rightarrow -\infty$ of 4. }
\vfill\eject

\psfig{figure=fig7.ps}
\fig{7}{Linearization of the dispersion relation around the Fermi
points yields the Luttinger model.  The cutoff $\Lambda$ is
introduced to avoid unphysical interactions between left and right
moving excitations.}
\vfill
\psfig{figure=fig8.ps}
\fig{8}{Diagrammatic representation of the four possible interactions
at zero momentum consistent with the symmetries of the problem.
Straight lines represent right moving and wavy lines left moving
Fermions.  Creation and annihilation operators are indicated by arrows
pointing respectively out of, or into, a vertex.}
\vfill\eject

\psfig{figure=fig9.ps}
\fig{9}{Phase diagram in the $(x_2 , y)$ plane.  The flows are identical
to those of the Kosterlitz--Thouless transition with $x_2$ playing the
role of the vortex fugacity, and $y$ taking the place of the reduced
temperature.  The shaded wedge represents the stable region within
which the spin--$1/2$ Luttinger--liquid description holds.  In the presence
of $SU(2)$ symmetry, the flows are restricted to the line $y=x_2$.}
\vfill
\psfig{figure=fig10.ps}
\fig{10}{The operators $O$ and $O^\dagger$ generating
 loops, pictured here for $p=2$.  For the general case, $O^\dagger$
and $O$ create and annihilate $2p$ steps respectively. The dashed line
indicates the closed path used to define the vortex configurations in
the displacement field description of Appendix C.}
\vfill\eject
\psfig{figure=fig11.ps}
\fig{11}{The addition of loops introduces this additional Fermion
vertex and its complex conjugate for $p=2$.}
\vfill
\psfig{figure=fig12.ps}
\fig{12}{Possible paths as a function of temperature through the
phase diagram for the case $p=2$. Path (a) is consistent with the
observations on $2 \times 1$ gold (110).  The branching of the paths
simply reflects our uncertainty whether the excitations in this region
are bound pairs or individual `wall' defects.  Path (b) represents a
direct transition between the FO and RD phases.  Path (c) is another
possibility for $2 \times 1$ systems which includes the new rough (RI)
phase.}
\vfill
\end